\documentclass{article}

\usepackage{amsmath}
\usepackage{amssymb}
\usepackage{program}
\usepackage{ifthen}
\usepackage{epsfig}
\usepackage{array}
\usepackage{epic}
\usepackage{color}
\usepackage{hyperref}
\usepackage{dirtree}
\usepackage{pifont}

\usepackage{algorithmicx,algpseudocode}

\newcolumntype{I}{!{\vrule width 1.5pt}}
\newlength\savedwidth

\newboolean{IsWide}
\setboolean{IsWide}{true}

\newcommand{\FlaPartition}[2]{
\ifthenelse{\boolean{IsWide}}{{\bf partition } \hspace{-1em} #1 \hspace{-1em} #2}
{{\bf partition } \+ \\ #1 \+ \\ #2 \- \-}
}

\newcommand{\FlaRepartition}[2]{
\ifthenelse{\boolean{IsWide}}{{\bf repartition } \hspace{-1em} #1 \hspace{-1em} #2}
{{\bf repartition } \+ \\ #1 \+ \\ #2 \- \-}
}

\newcommand{\FlaContinue}[1]{
\ifthenelse{\boolean{IsWide}}{{\bf continue with } #1
}
{{\bf continue with } \+ \\ #1 \-
}
}

\newcommand{\blocksize}{1}

\newcommand{\repartitionings}{
\begin{minipage}[t]{2in}
\ \\
\ \\
\ \\
\end{minipage}
}

\newcommand{\repartitionsizes}{ \hspace{ 1.25in} }

\newcommand{\WSrepartition}{
\begin{minipage}[t]{2in}
\ifthenelse{ \equal{\blocksize}{1} }{}
{%
\ifthenelse{ \equal{\blocksize}{2} }{~}
{\bf Determine block size $ \blocksize $} \\
}
{\bf Repartition}
\begin{tabbing}
in \= in \= \+ \kill
\repartitionings \+ \\
{\bf where } \hspace*{-2ex} \repartitionsizes 
\end{tabbing}
\end{minipage}
}

\newcommand{\WSrepartitionNarrow}{
\begin{minipage}[t]{2.15in}
\ifthenelse{ \equal{\blocksize}{1} }{}
{%
\ifthenelse{ \equal{\blocksize}{blank} }{~}
{\bf Determine block size $ \blocksize $} \\
}
{\bf Repartition}
\begin{tabbing}
i \= i \= \+ \kill
\repartitionings \+ \\
{\bf where } \hspace*{-2ex} \repartitionsizes 
\end{tabbing}
\end{minipage}
}

\newcommand{\Gemm}{\mbox{\sc Gemm}}

\newcommand{\NoShow}[1]{}
\newcommand{\myhref}[2]{\href{#1}{\ding{42} #2}}

\newcommand{\Jc}{{\cal J}_c}
\newcommand{\Pc}{{\cal P}_c}
\newcommand{\Ic}{{\cal I}_c}
\newcommand{\Jr}{{\cal J}_r}

\newcommand{\Ir}{{\cal I}_r}

\begin{document}

\title{
BLISlab: A Sandbox for Optimizing GEMM
\\[0.2in]
\large FLAME Working Note \#80
\\[0.2in]
\normalsize Use appropriate PDF viewer so that hyperlinks (denoted with \ding{42}) work
}

\author{
Jianyu Huang
\and
Robert A. van de Geijn
}

\date{August 31, 2016}

\maketitle

\begin{abstract}
Matrix-matrix multiplication is a fundamental operation of great importance to scientific computing and, increasingly, machine learning.  
It is a simple enough concept to be introduced in a typical high school algebra course yet in practice important enough that its implementation on computers continues to be an active research topic.
This note describes a set of exercises that use this operation to illustrate how high performance can be attained on modern CPUs with hierarchical memories (multiple caches).  It does so by building on the insights that underly the BLAS-like Library Instantiation Software (BLIS) framework by exposing a simplified ``sandbox'' that mimics the implementation in BLIS.
As such, it also becomes a vehicle for the ``crowd sourcing'' of the optimization of BLIS.
We call this set of exercises ``\myhref{https://github.com/flame/blislab}{BLISlab}''\footnote{https://github.com/flame/blislab}.

\end{abstract}

\section{Introduction}
\label{sec:introduction}

Matrix-matrix multiplication (\Gemm) is frequently used as a simple example with which to raise awareness of how to optimize code on modern processors.  The reason is that the operation is simple to describe, challenging to fully optimize, and of practical importance.  In this document, we walk the reader through the techniques that underly the currently fastest implementations for CPU architectures.

\subsection{Basic Linear Algebra Subprograms (BLAS)}

The Basic Linear Algebra Subprograms (BLAS)~\cite{BLAS1,BLAS2,BLAS3,BLIS-Encycl} form an interface for a set of linear algebra operations upon which higher level linear algebra libraries, such at LAPACK~\cite{LAPACK3} and {\tt libflame}~\cite{libflame_ref}, are built.  The idea is that if someone optimizes the BLAS for a given architecture, then all applications and libraries that are written in terms of calls to the BLAS will benefit from such optimizations.  

The BLAS are divided into three sets: the level-1 BLAS (vector-vector operations), the level-2 BLAS (matrix-vector operations), and the level-3 BLAS (matrix-matrix operations).
The last set benefits from the fact that, if all matrix operands are $ n \times n$ in size, $ O( n^3 ) $ floating point operations are performed with $ O( n^2 ) $ data so that the cost of moving data between memory layers (main memory, the caches, and the registers) can be amortized over many computations.  As a result, high performance can in principle be achieved if these operations are carefully implemented.

\subsection{Matrix-matrix multiplication}

In particular, \Gemm\ with double precision floating point numbers is supported by the BLAS with the (Fortran) call
\begin{verbatim}

      dgemm( transa, transb, m, n, k alpha, A, lda, B, ldb, beta, C, ldc )
      
\end{verbatim}
which, by appropriately choosing {\tt transa} and {\tt transb}, 
computes 
\[
C := \alpha A B + \beta C; \quad
C := \alpha A^T B + \beta C; \quad
C := \alpha A B^T + \beta C; \quad \mbox{or }
C := \alpha A^T B^T + \beta C.
\]
Here $ C $ is $ m \times n $ and $ k $ is the ``third dimension''.  The parameters {\tt dla}, {\tt dlb}, and {\tt dlc} are explained later in this document.

\NoShow{
	Of importance in this call is the concept of a {\em leading dimension}.
Consider the array of numbers
\[
\begin{array}{r r r}
 1.1 & \color{red} 1.2 &\color{red}1.3 \\
 2.1 & \color{red} 2.2 &\color{red}2.3 \\
3.1 &3.2 &3.3 \\
\end{array}
\]
which, when column major order is used, would be stored in memory contiguously as
\[
\begin{array}{r}
1.1 \\
2.1 \\
3.1 \\
\color{red} 1.2 \\
\color{red} 2.2 \\
3.2 \\
\color{red} 1.3 \\
\color{red} 2.3 \\
3.3 
\end{array}
\]
Now consider the matrix $ 
A = \left(\begin{array}{r r}
 \color{red} 1.2 &\color{red}1.3 \\
 \color{red} 2.2 &\color{red}2.3 \\
\end{array} \right)$.
In the call to {\tt dgemm} this matrix could be passed to the routine by choosing
{\tt k=2}, {\tt n=2}, {\tt A} equal to the address of where $ 1.2 $ is stored in memory, and {\tt lda=3}.
The reason why {\tt lda=3} is that the matrix is stored as a subarray of an array that itself has $ 3 $ entries per column, which must be specified for the routine to understand how to access the matrix in memory.  Another way of thinking about this is that accessing elements of a row of $ A $ one must "stride" through memory with a stride of $ 3 $ elements.
}

In our exercises, we consider the simplified version of \Gemm,
\[
C := A B + C,
\]
where $ C $ is $ m \times n $, 
$ A $ is $ m \times k $, and $ B $ is $ k \times n $.
If one understands how to optimize this particular case of {\tt dgemm}, then one can easily extend this knowledge to all level-3 BLAS functionality.

\subsection{High-performance implementation}

The intricacies of high-performance implementations are such that implementation of the BLAS in general and \Gemm\ in particular was 
often relegated to 
unsung experts who develop numerical libraries for the hardware vendors, for example as part of IBM's ESSL, Intel's MKL, Cray's LibSci, and AMD's ACML libraries.  These libraries were typically written (at least partially) in assembly code and highly specialized for a specific processor.
  
A key paper~\cite{IBM:P2} showed how an ``algorithms and architectures" approach to hand-in-hand designing architectures, compilers, and algorithms allowed 
BLAS to be written in a high level language (Fortan) for the IBM Power architectures and explained the intricacies of achieving high performance on those processors.
The Portable High Performance ANSI C (PHiPAC)~\cite{PHiPAC97} project subsequently provided guidelines for writing high-performance code in C and suggested how to autogenerate and tune \Gemm\ written this way.  The Automatically Tuned Linear Algebra Software (ATLAS)~\cite{ATLAS,ATLAS_journal} built upon these insights and made autotuning and autogeneration of BLAS libraries mainstream.

As part of this document we discuss more recent papers on the subject, including the paper that introduced the Goto approach to implementing \Gemm~\cite{Goto:2008:AHP} and the BLIS refactoring of that approach~\cite{BLIS1}, as well as other papers that are of more direct relevance.  

\subsection{Other similar exercises}

There are others who have put together exercises based on {\Gemm}.
Recent efforts relevant to this paper are
\myhref{http://apfel.mathematik.uni-ulm.de/~lehn/sghpc/gemm/index.html}{GEMM: From Pure C to SSE Optimized Micro Kernels}
by Michael Lehn at Ulm University and  a wiki on 
\myhref{https://github.com/flame/how-to-optimize-gemm/wiki}{Optimizing Gemm} that we ourselves put together.

\subsection{We need you!}

The purpose of this paper is to guide you towards high-performance
implementations of \Gemm.  Our ulterior motive is that our BLIS framework for implementing BLAS requires a so-called micro-kernel to be highly optimized for various CPUs.  In teaching you the basic techniques, we are hoping to identify  ``The One'' who will contribute the best micro-kernel.
Think of it as our version of ``HPC's Got Talent''.
Although  we focus in our description on optimization for the Intel Haswell architecture, the setup can be easily modified to instead help you (and us) optimize for other CPUs.  Indeed, BLIS itself supports architectures that include AMD and Intel's x86 processors, IBM's Power processors, ARM processors, and Texas Instrument DSP processors~\cite{BLIS2,BLIS3,BLIS-TI}.

\section{Step 1: The Basics}

\subsection{Simple matrix-matrix multiplication}

In our discussions, we will consider the computation
$$C:=AB + C$$
where $A$, $B$, and $C$ are $m\times k$, $k\times n$, $m\times n$  matrices, respectively. 
Letting 
{\footnotesize%
\[
A = 
\left( \begin{array}{c c c c}
a_{0,0} & \cdots & a_{0,k-1} \\
\vdots &  & \vdots \\
a_{m-1,0} & \cdots & a_{m-1,k-1} \\
\end{array}
\right),
B = 
\left( \begin{array}{c c c c}
b_{0,0} & \cdots & b_{0,n-1} \\
\vdots &  & \vdots \\
b_{k-1,0} & \cdots & b_{k-1,n-1} \\
\end{array}
\right), \mbox{~and~}
C = 
\left( \begin{array}{c c c c}
c_{0,0} & \cdots & c_{0,n-1} \\
\vdots &  & \vdots \\
c_{m-1,0} & \cdots & c_{m-1,n-1} \\
\end{array}
\right)
\]%
}
$ C := A B + C $ computes
\[
c_{i,j} := \sum_{p=0}^{k-1} a_{i,p} b_{p,j} +
c_{i,j}.
\]
If $ A $, $ B $, and $ C $ are stored  in two-dimensional arrays {\tt
  A}, {\tt B}, and {\tt C}, 
the following pseudocode computes $ C := A B + C $:

\vspace{0.1in}
\begin{center}
\begin{minipage}{4in}
\begin{verbatim}
for i=0:m-1
   for j=0:n-1
      for p=0:k-1
         C( i,j ) := A( i,p ) * B( p,j ) + C( i,j )
      endfor
   endfor
endfor
\end{verbatim}
\end{minipage}
\end{center}
\vspace{0.1in}
Counting a multiply and an add separately,
the computation requires $ 2 m n k $  floating point operations (flops).

\subsection{Setup}

To let you efficiently learn about how to efficiently compute, you
start your project with much of the infrastructure in place.  We have
structured the subdirectory, {\tt step1}, somewhat like a project that
implements a real library might.   This may be overkill for our
purposes, but how to structure a software project is a useful skill to learn.

\begin{figure}[tb!]
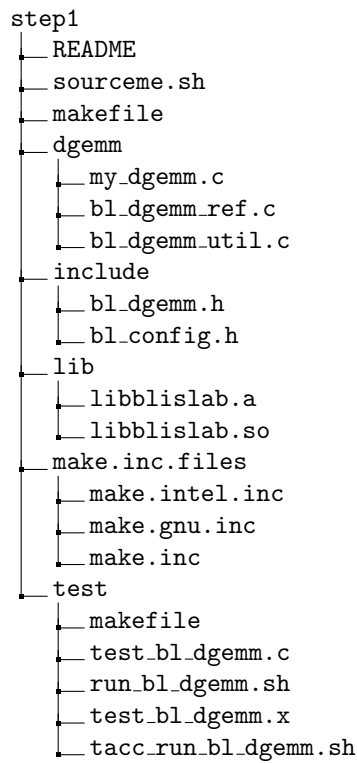

	\begin{center}
\begin{minipage}{4in}
	\dirtree{%
		.1 step1.
		.2 README 
.
		.2 sourceme.sh 
.
		.2 makefile 
.
		.2 dgemm 
.
		.3 my\_dgemm.c 
.
		.3 bl\_dgemm\_ref.c 
.
		.3 bl\_dgemm\_util.c 
.
		.2 include 
.
		.3 bl\_dgemm.h 
.
		.3 bl\_config.h 
.
		.2 lib 
.
        .3 libblislab.a
.
        .3 libblislab.so
.
		.2 make.inc.files
.
		.3 make.intel.inc 
.
		.3 make.gnu.inc 
.
		.3 make.inc 
.
		.2 test 
.
		.3 makefile 
.
		.3 test\_bl\_dgemm.c 
.
		.3 run\_bl\_dgemm.sh 
.
		.3 test\_bl\_dgemm.x 
.       
		.3 tacc\_run\_bl\_dgemm.sh 
.	
	}
\end{minipage}
\end{center}
\caption{Structure of directory {\tt step1}.}
\label{fig:DirStep1}
\end{figure}

Consider Figure~\ref{fig:DirStep1}, which illustrates the directory
structure for subdirectory {\tt step1}:
\begin{description}
\item[{\tt README}]
Is a file that describes the contents of the directory and how to compile and execute the code.
\item[{\tt sourceme.sh}]
Is a file that configures the environment variables.  In that file
\begin{description}
\item[{\tt BLISLAB\_USE\_INTEL}] sets whether you use the Intel compiler ({\tt true}) or the GNU compiler ({\tt false}).
\item[{\tt BLISLAB\_USE\_BLAS}] indicates whether your reference {\tt dgemm} employs an external {\tt BLAS} library implementation ({\tt true} if you have such a BLAS library installed on your machine), or the simple triple loops implementation ({\tt false}).
\item[{\tt COMPILER\_OPT\_LEVEL}] sets the optimization level for your GNU or Intel compiler ({\tt O0}, {\tt O1}, {\tt O2}, {\tt O3}).  (Notice that, for example, {\tt O3} consists of the capital letter "O" and the number "3".)
\item[{\tt OMP\_NUM\_THREADS} and {\tt BLISLAB\_IC\_NT}] sets the number of threads used for parallel version of your code. For Step 1, you set them both to {\tt 1}.
\end{description}
\item[{\tt dgemm}]
Is the subdirectory where the routines that implement {\tt dgemm} exist.
In it
\begin{description}
\item[{\tt bl\_dgemm\_ref.c}] contains the routine {\tt dgemm\_ref} that
  is a simple implementation of {\tt dgemm} that you will use to check the
  correctness of your implementations, if {\tt BLISLAB\_USE\_BLAS = false}.
\item[{\tt my\_dgemm.c}] contains the routine {\tt dgemm} that
  that initially is a simple implementation of {\tt dgemm} and that you will optimize as part of the first step on your way to mastering how to optimize {\tt gemm}.
\item[{\tt bl\_dgemm\_util.c}] contains utility routines that will later come in handy.
\end{description}
\item[{\tt include}]
This directory contains 
include files with various macro definitions and 
other header information.
\item[{\tt lib}] This directory will hold libraries generated by your implemented source files ({\tt libblislab.so} and {\tt libblislab.a}). You can also install a reference library (e.g. OpenBLAS) in this directory to
compare your performance.
\item[{\tt test}] This directory contains ``test drivers'' and correctness/performance checking scripts for the various implementations.
\begin{description}
\item[{\tt test\_bl\_dgemm.c}] contains the ``test driver'' for testing routine {\tt bl\_dgemm}. 
\item[{\tt test\_bl\_dgemm.x}] is the executable file for {\tt test\_bl\_dgemm.c}. 
\item[{\tt run\_bl\_dgemm.sh}] contains a bash script to collect performance results.
\item[{\tt tacc\_run\_bl\_dgemm.sh}] contains a {\tt SLURM} script for you to (optionally) submit the job to the Texas Advanced Computing Center (TACC) machines if you have an account there.
\end{description}
\end{description}

\subsection{Getting started}

\begin{figure}[tb!]
	\begin{center}
		\begin{minipage}{4.5in}
        \begin{verbatim}
        for ( i = 0; i < m; i ++ ) {                 // 2-th loop
          for ( j = 0; j < n; j ++ ) {               // 1-th loop
            for ( p = 0; p < k; p ++ ) {             // 0-th loop
              C( i, j ) += A( i, p ) * B( p, j );
            }                                        // End 0-th loop
          }                                          // End 1-th loop
        }                                            // End 2-th loop
        \end{verbatim}
		\end{minipage}
	\end{center}
	\caption{Simple implementation of \Gemm.}
	\label{fig:threeloops}
\end{figure}

What we want you to do is to start with the implementation in {\tt my\_dgemm.c} and optimize it by applying various standard optimization techniques.
The initial implementation in that file is the straight-forward implementation with the three loops given in Figure~\ref{fig:threeloops}.
The first thing to notice is how two-dimensional arrays are mapped to memory in so-called {\em column-major order}.  The reason for this choice is that the original BLAS assumed column-major storage of arrays because the interface was for Fortran users first.  
Examining
\begin{verbatim}

          C( i, j ) += A( i, p ) * B( p, j );
          
\end{verbatim}
we notice that, each operand is a {\tt MACRO}. Consider early in that file
\begin{verbatim}

   #define C( i, j ) C[ (j)*ldc + (i) ]
   
\end{verbatim}
The linear array at address $ C $ is used to store elements $ C_{i,j} $ so that the $ i,j $ element is mapped to location {\tt j * ldc + i}.  The way to view this is that the columns of $ C $ are each stored contiguously.  However, think of matrix $ C $ as embedded in a larger array that has {\tt ldc} rows so that accessing a row means going through array {\tt C} with stride {\tt ldc}.  The term {\em leading dimension} of two-dimensional array {\tt C} is typically used to refer to the row dimension of this larger array, hence the variable {\tt ldc} (\underline{l}eading \underline{d}imension of \underline{\tt C}).
This is illustrated for all three matrices in
the following figure:
\begin{center}
	\includegraphics[width=5in]{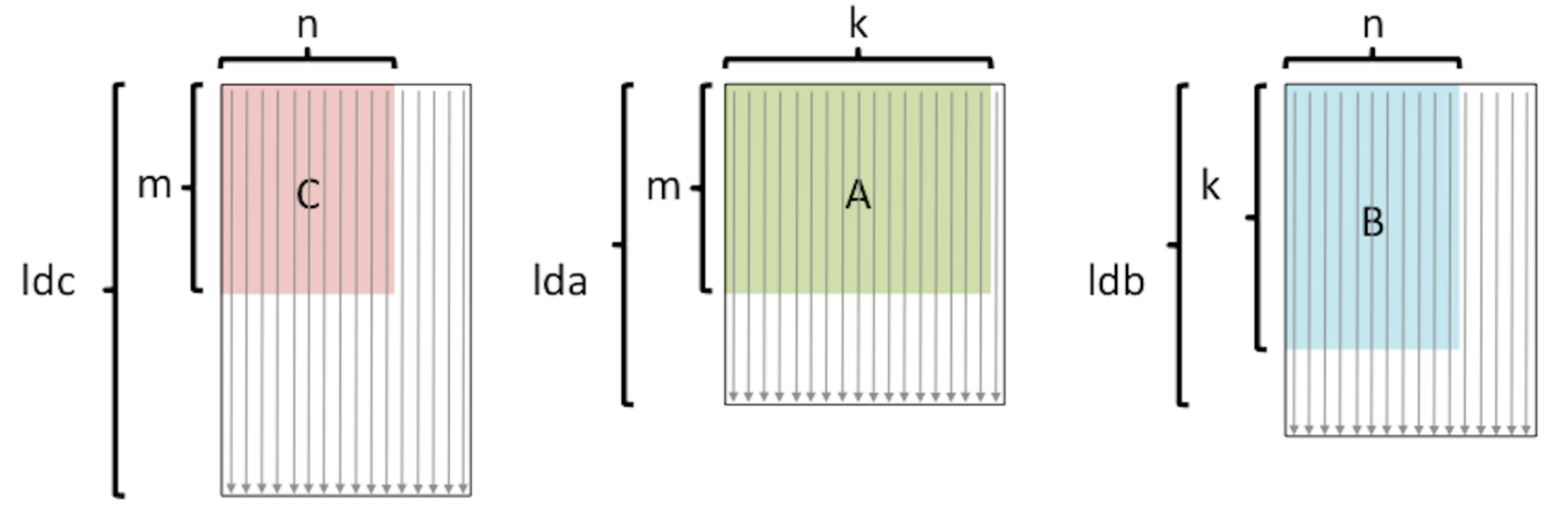}
\end{center}
in which the arrows are meant to indicate that columns are stored contiguously.

\subsubsection{Configure the default implementation}
By default, the exercise compiles and links with Intel's {\tt icc} compiler, which will apply compiler optimizations (level {\tt O3}) to the code.
You need to set the environment variable by executing:
\begin{verbatim}

   $ source sourceme.sh
   
\end{verbatim}
in the terminal, and you will see the output:
\begin{verbatim}

   BLISLAB_USE_INTEL = true
   COMPILER_OPT_LEVEL = O3

\end{verbatim}

\subsubsection{Compile, execute and collect results}
\label{sec:mkl}

	{\color{black} If you do not have access to Intel's compiler ({\tt icc}), then read Subsections~\ref{sec:mkl} and~~\ref{sec:graph}, and continue with Subsection~\ref{sec:gcc}.}

\vspace{0.1in}
\noindent
You can compile, execute your code and collect the performance result by executing
\begin{verbatim}

   make clean
   make
   cd test
   ./run_bl_dgemm.sh

\end{verbatim}
in subdirectory{\tt step1}. You will see the performance result output:
\begin{verbatim}

   run_step1_st=[
   %m     %k   %n   %MY_GFLOPS %REF_GFLOPS
   16    16    16  0.82  2.15
   32    32    32  0.74  5.50
   48    48    48  0.85  5.66
   ......
   ];

\end{verbatim}
You can change the sampling block size in {\tt run\_bl\_dgemm.sh}.
Notice that if you have errors in your code, these will be reported as, for example, 
\begin{verbatim}

C[ 0 ][ 0 ] != C_ref, 1.253000E+00, 2.253000E+00

\end{verbatim}

\subsubsection{Draw the performance graph}
\label{sec:graph}

Finally, you can use {\tt MATLAB} to draw your performance graph with our scripts. In {\tt misc/results} subdirectory, after executing 
\begin{verbatim}

   ./collect_result_step1

\end{verbatim}
you will get a {\tt MATLAB} file ``{\tt step1\_result.m}'', with the performance results. You can then execute 
\begin{verbatim}

   bl_dgemm_plot.m

\end{verbatim}
in {\tt MATLAB}, which will then generate the performance graph. 

\subsubsection{Change to the GNU compiler}

Since we want you to explicitly learn about what kind of tricks lead to high performance, and because some of you may not have access to the Intel compiler, you should next change to using the GNU C compiler.
For this, you must edit {\tt sourceme.sh}: 
\begin{verbatim}

   BLISLAB_USE_INTEL=false
   
\end{verbatim}
Then, similar to the default setting, you need to set the environment variable by executing:
\begin{verbatim}

   $ source sourceme.sh
   
\end{verbatim}
in the terminal, and you will observe:
\begin{verbatim}

   BLISLAB_USE_INTEL = false
   COMPILER_OPT_LEVEL = O3
   
\end{verbatim}

\subsubsection{Turn off optimization}
\label{sec:gcc}

Next, we want you to turn off the optimization performed by the compiler.  This serves three purposes: first, it means you are going to have to explicitly perform optimizations, which will allow you to learn about how architectures and algorithms interact.  Second, it may very well be that the optimizing compiler will try to ``undo'' what you are explicitly trying to accomplish.  Third, the more tricks you build into your code, the harder it gets for the compiler to figure out how to optimize.

You need first edit {\tt sourceme.sh}: 
\begin{verbatim}

   COMPILER_OPT_LEVEL = O0

\end{verbatim}
Then, similar to the default setting, you need to set the environment variable by executing:
\begin{verbatim}

   $ source sourceme.sh
   
\end{verbatim}
in the terminal, and you will see the output:
\begin{verbatim}

  BLISLAB_USE_INTEL = false
  COMPILER_OPT_LEVEL = O0
  
\end{verbatim}

\subsubsection{(\textbf{Optional}) Use optimized BLAS library as reference implementation}

By default, your reference \Gemm\ implementation is a very slow triple-loop implementation.
If you have a {\tt BLAS} library installed on your test machine, you can adopt the {\tt dgemm} from that library as your reference implementation by setting:
\begin{verbatim}

   BLISLAB_USE_BLAS=true
   
\end{verbatim}
in {\tt sourceme.sh}.
If you use Intel compiler, you don't need to explicitly specify the path of MKL. However, if you use GNU compiler, you need to specify the path of your BLAS library.   For example, you may want to install our BLIS library from 
\myhref{https://github.com/flame/blis}{\url{https://github.com/flame/blis}}
in directory {\tt /home/lib/blis} and in {\tt sourceme.sh} set
\begin{verbatim}

   BLAS_DIR=/home/lib/blis

\end{verbatim}
After executing {\tt \$ source sourceme.sh}, you will observe:
\begin{verbatim}

  BLISLAB_USE_BLAS = true
  BLAS_DIR = /home/lib/blis

\end{verbatim}
and now performance and accuracy comparisons of your implementation will be against this optimized library routine.


\subsection{Basic techniques}

In this subsection we describe some basic tricks of the trade.

\subsubsection{Using pointers}

Now that optimization is turned off, the computation of the address where an element of a matrix exists is explicitly exposed.  (An optimizing compiler would get rid of this overhead.)
What you will want to do is to change the implementation in {\tt my\_gemm.c} so that it instead uses pointers.
Before you do so, you may want to back up the original {\tt my\_gemm.c} in case you need to restart from scratch.  Indeed, at each step you may want to back up in a separate file the previous implementations.

Here is the basic idea.  Let's say we want to set 
all elements of $C $ to zero.  A basic loop, patterned after what you found in {\tt my\_gemm.c} might look like
\begin{verbatim}

   for ( i = 0; i < m; i ++ ) {                   
      for ( j = 0; j < n; j ++ ) {                
         C( i, j ) = 0.0;
      }                                           
   }                                              
\end{verbatim}
Using pointers, we might implement this as
\begin{verbatim}

   double *cp;

   for ( j = 0; j < n; j ++ ) {
      cp = &C[ j * ldc ];           // point cp to top of jth column      
      for ( i = 0; i < m; i ++ ) {                
         *cp++ = 0.0;                // set the element that cp points to to zero and 
                                     // advance the pointer.
      }                                      
   }                 
   
\end{verbatim}
Notice that we purposely exchanged the order of the loops so that advancing the pointer takes us down the columns of $ C $.

\subsubsection{Loop unrolling}

Updating loop index {\tt i} and the pointer {\tt cp} every time through the inner loop creates considerable overhead.  For this reason, a compiler will perform {\em loop unrolling}.  Using an unrolling factor of four, our simple loop for setting {\tt C} to zero becomes
 \begin{verbatim}
 
    double *cp;
 
    for ( j = 0; j < n; j ++ ) {
    cp = &C[ j * ldc ];           
       for ( i = 0; i < m; i += 4 ) {                
          *(cp+0) = 0.0; 
          *(cp+1) = 0.0; 
          *(cp+2) = 0.0; 
          *(cp+3) = 0.0;
          cp += 4;                
          }                                           
       }  
              
 \end{verbatim}
Importantly:
\begin{itemize}
	\item {\tt i} and {\tt cp} are now only updates once every four iterations.
	\item {\tt *(cp+0)} uses a machine instruction known as {\em indirect addressing} that is much more efficient than if one computed with {\tt *(cp+k)} where $ k $ is a variable.  
	\item
	When data it brought in for memory into cache, it is brought in a cache line of 64 bytes at a time.  This means that accessing contiguous data in chunks of 64 bytes reduces the cost of memory movement between the memory layers.
\end{itemize}
Notice that when you unroll, you may have to deal with a ``fringe'' if, in this case, {\tt m} is not a multiple of four.    For the sake of this exercise, you need not worry about this fringe {\em as long as you pick your sampling block size wisely}, as reiterated in Section~\ref{sec:two:mission}.

\subsubsection{Register variables}

Notice that computation can only happen if data is stored in registers.  A compiler will automatically transform code so that the intermediate steps that place certain data in registers is inserted.  One can give a hint to the compiler that it would be good to keep certain data in registers as illustrated in the following somewhat contrived example:
\begin{verbatim}

   double *cp;

   for ( j = 0; j < n; j ++ ) {
      cp = &C[ j * ldc ];           
      for ( i = 0; i < m; i += 4 ) {
         register double c0=0.0, c1=0.0, c2=0.0, c3=0.0;                
         *(cp+0) = c0; 
         *(cp+1) = c1; 
         *(cp+2) = c2; 
         *(cp+3) = c3;
         cp += 4;                
      }                                           
   }       
\end{verbatim}

\subsection{A modest first goal}
\label{sec:two:mission}

We now ask you to employ the techniques discussed above to optimize {\tt my\_dgemm}.  For now, just worry about trying to attain better performance for smallish matrices.  In particular, consider the following picture:
\begin{center}
	\includegraphics[width=3in]{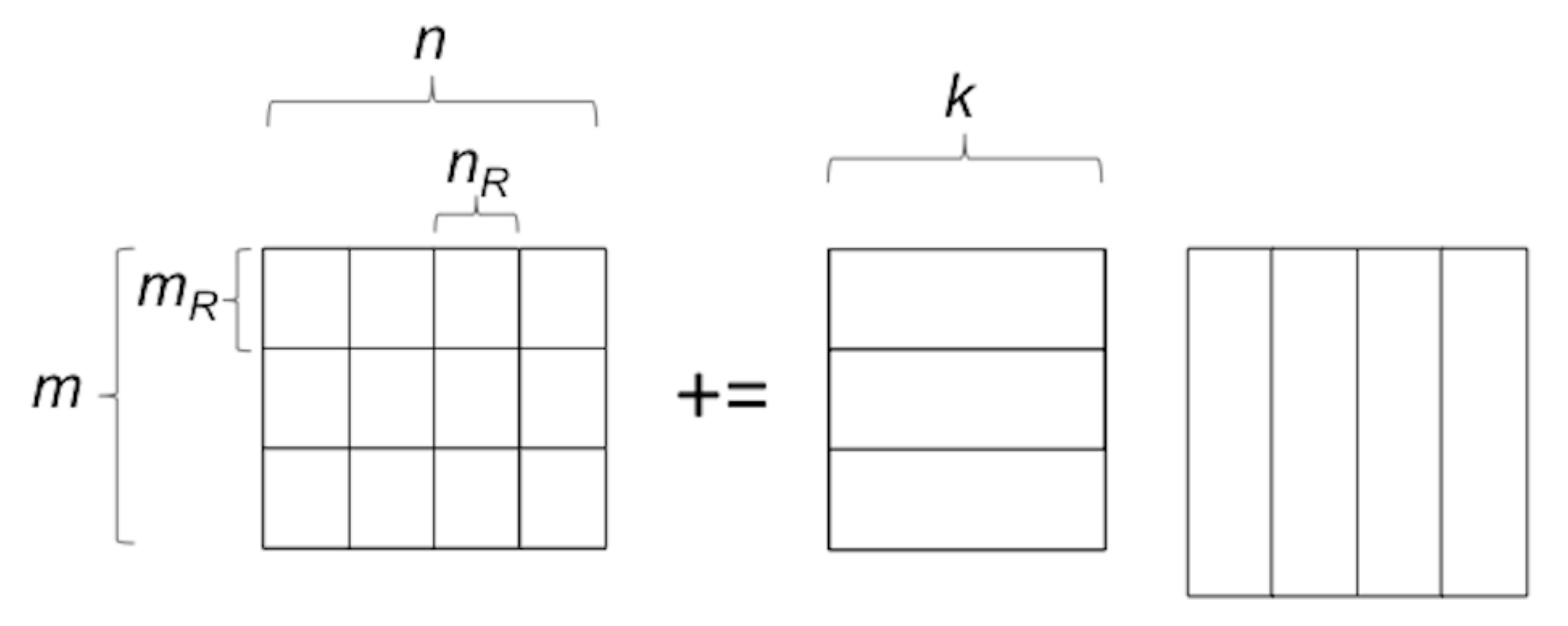}
\end{center}
What we want you to do is to write your code so that $ m_R \times n_R $ blocks of $ C $ are kept in registers.  
You get to choose $ m_R $ and $ n_R $, but you will want to update file {\tt include/bl\_config.h} with those choices.
This ensures that the test driver only tries problem sizes that are multiples of these block sizes, so you don't have to worry about ``fringe''.


You will notice that even for smallish matrices that can fit in one of the cache memories, your implementation performs (much) worse than the implementations that are part of MKL or other optimized BLAS library that you may have installed.  
The reason is that the compiler is not using the fastest instructions for floating point arithmetic.  These can be accessed either by using {\em vector intrinsic functions}, which allows you to explicitly utilize them from C, or by coding in assemply code.  For now, let's not yet go there.  We will talk more about this in Step 3.

\section{Step 2: Blocking}

\subsection{Poorman's BLAS}

Step 1 of this exercise makes you realize that
with the advent of cache-based architectures, high-performance implementation of \Gemm\ necessitated careful attention to 
the amortization of the cost of data movement between memory layers and computation with that data.
To keep this manageable, it helps to realize that only a ``kernel'' that performs a  matrix-matrix multiplication with relatively small matrices needs to be highly optimized, since
computation with larger matrices can be blocked to then use such a kernel without an adverse impact on overall performance.  This 
insight was explicitly advocated in~\cite{poorman_journal}
\begin{quote}
	Bo{\aa}gstr\"{o}m , Per Ling, Charles Van Loan.
	\myhref{http://dl.acm.org/citation.cfm?id=292412&CFID=766560304&CFTOKEN=45870613}{GEMM-based level 3 BLAS: high-performance model implementations and performance evaluation benchmark.}
	ACM Transactions on Mathematical Software (TOMS). 
	Volume 24 Issue 3, p.268-302, Sept. 1998.
\end{quote}
This is sometimes referred to as "poorman's BLAS" in the sense that if one could only afford to optimize matrix-matrix multiplication (with submatrices), then one could build \Gemm, and other important matrix-matrix operations known as the level-3 BLAS, in terms of this.  What we will see later is that actually in general this is a good idea, for the sake of modularity as well as for performance.

In the last section you already saw an example of blocking.  

\subsection{Blocked matrix-matrix multiplication}

Key to blocking \Gemm\ to take advantage of the hierarchical memory of
a processor is understanding how to compute $ C := A B + C $ when
these matrices have been blocked.  Partition
{\footnotesize%
\[
A = 
\left( \begin{array}{c c c c}
A_{0,0} & \cdots & A_{0,K-1} \\
\vdots &  & \vdots \\
A_{M-1,0} & \cdots & A_{M-1,K-1} \\
\end{array}
\right),
B = 
\left( \begin{array}{c c c c}
B_{0,0} & \cdots & B_{0,N-1} \\
\vdots &  & \vdots \\
B_{K-1,0} & \cdots & B_{K-1,N-1} \\
\end{array}
\right), \mbox{~and~}
C = 
\left( \begin{array}{c c c c}
C_{0,0} & \cdots & C_{0,N-1} \\
\vdots &  & \vdots \\
C_{M-1,0} & \cdots & C_{M-1,N-1} \\
\end{array}
\right).
\]%
}
where $ C_{i,j} $ is $ m_i \times n_j
$, $ A_{i,p} $ is $ m_i \times k_p
$, and $ B_{p,j} $ is $ k_p \times n_j
$.
Then
\[
C_{i,j} := \sum_{p=0}^{K-1} A_{i,p} B_{p,j} + C_{i,j}.
\]

\subsection{Your mission, if you choose to accept it}
We now ask you to implement the blocked matrix-matrix multiplication in {\tt my\_dgemm}.
Specifically, for small matrices you achieve better performance than for larger matrices because the smaller matrices fit in cache.  Block the matrices into submatrices of the size for which you do attain higher performance, and you will see that the resulting implementation can maintain the better performance even for larger matrices.

\section{Step 3: Blocking for Multiple Levels of Cache}

\subsection{The Goto Approach to Implementing {\sc gemm}}
\label{sec:BLIS}

\begin{figure}[tb!]
	\begin{center}
		\begin{minipage}{3in}
			\mbox{\includegraphics[width=3.0in]{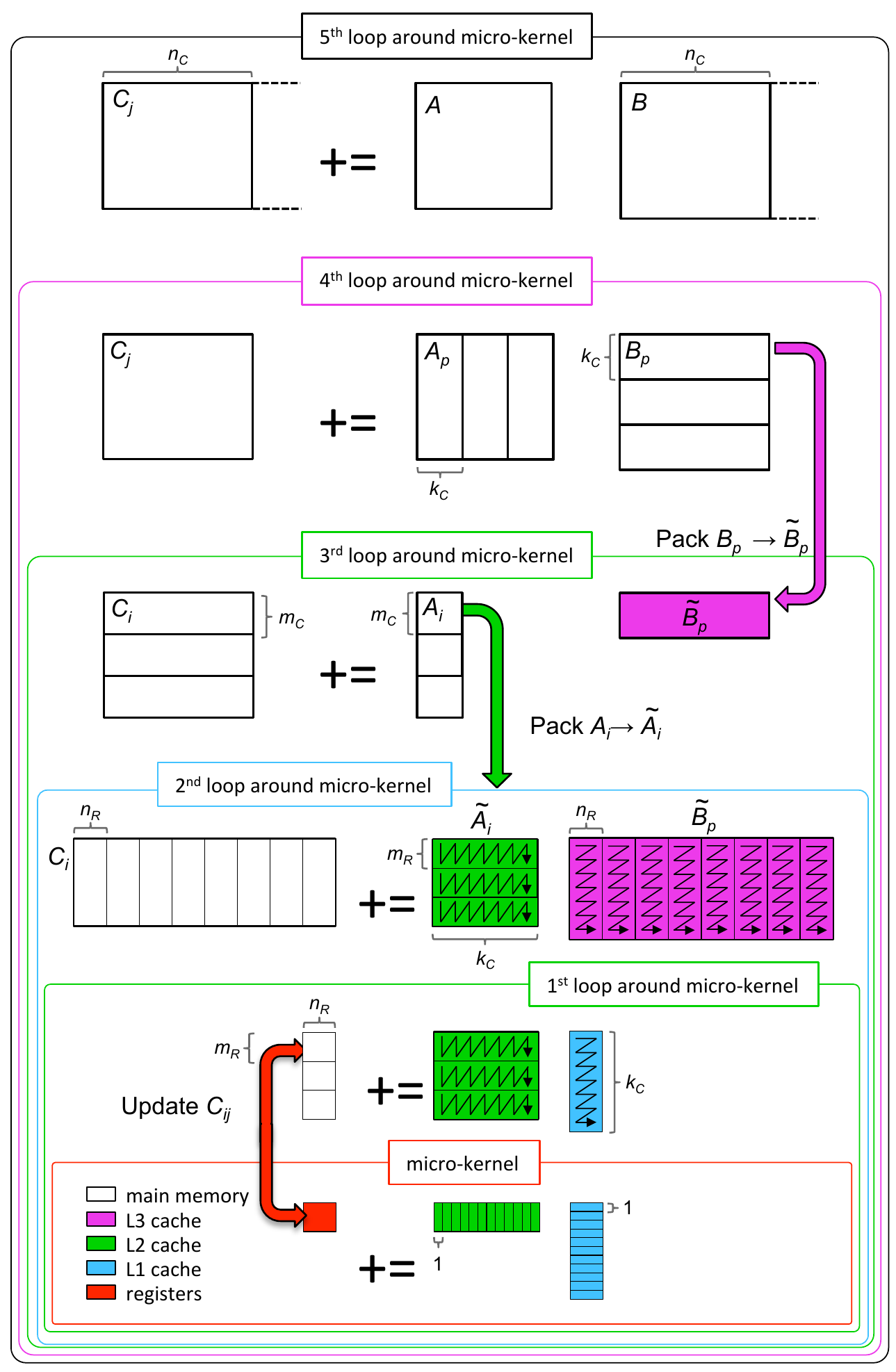}}
		\end{minipage}
		~~~
		\begin{minipage}[t]{3in}
			\footnotesize  
			\mbox{\begin{tabular}{l@{\hspace{6pt}}l@{}}
Loop 5 &{\bf for} $j_c\!=\! 0:n\!-\!1$ {\bf steps of} $n_c$ \\
& \hspace{2ex}  $\Jc\!=\! j_c:j_c\!+\!n_c\!-\!1$\\
Loop 4 & \hspace{2ex}  {\bf for} $p_c \!=\! 0:k\!-\!1$ {\bf steps of}
         $k_c$ \\
& \hspace{4ex} $\Pc \!=\! p_c:p_c\!+\!k_c\!-\!1$\\
&\hspace{4ex}           \textcolor{black}{$B(\Pc,\Jc)$} $\rightarrow \textcolor{black}{B_c}$ ~~~~~~~~~~ // Pack into $B_c$\\
Loop 3 & \hspace{4ex}           {\bf for} $i_c \!=\! 0:m\!-\!1$ {\bf
         steps of} $m_c$ \\
& \hspace{6ex} $\Ic \!=\! i_c:i_c\!+\!m_c\!-\!1$\\
&\hspace{6ex}                     \textcolor{black}{$A(\Ic,\Pc)$} $\rightarrow \textcolor{black}{A_c}$ ~~~~~~~~~~ // Pack into $A_c$ \\ \cline{2-2} 
& \hspace{6ex} // Macro-kernel\\ 
Loop 2 &\hspace{6ex} {\bf for} $j_r \!=\! 0:n_c\!-\!1$ {\bf steps of}
         $n_r$ \\
& \hspace{8ex}  $\Jr \!=\! j_r:j_r\!\!+\!\!n_r\!-\!1$  \\
Loop 1&\hspace{8ex}   {\bf for} $i_r \!=\! 0:m_c\!-\!1$ {\bf  steps
        of} $m_r$ \\
& \hspace{10ex}  $\Ir \!=\! i_r:i_r\!+\!m_r\!-\!1$\\
\cline{2-2} 
&\hspace{10ex} // Micro-kernel \\
Loop 0&\hspace{10ex}          {\bf for} $k_r \!=\! 0:k_c\!-\!1$ 
\\
&\hspace{16ex}             \textcolor{black}{$C_c(\Ir,\Jr)$}  \\
& \hspace{21ex}~$\mathrel{\!+\!}=$~\textcolor{black}{$A_c(\Ir,k_r)$} 
~\textcolor{black}{$B_c(k_r,\Jr)$} \\
&\hspace{10ex} {\bf endfor}\\
\cline{2-2} 
&\hspace{8ex} {\bf endfor}\\
&\hspace{4ex} {\bf endfor}\\
\cline{2-2} 
&\hspace{2ex} {\bf endfor}\\ 
&{\bf endfor}\\ 
\end{tabular}
  }
		\end{minipage}
	\end{center}
	\caption{Left: The GotoBLAS algorithm for matrix-matrix multiplication as  
		refactored in BLIS.  Right: the same algorithm, but expressed as  
		loops.}
	\label{fig:blis_gemm}
\end{figure}

Around 2000, Kazushige Goto revolutionized how \Gemm\ is implemented on current CPUs with his techniques
that were first published in the paper~\cite{Goto:2008:AHP}
\begin{quote}
	Kazushige Goto, Robert A. van de Geijn.
	\myhref{http://dl.acm.org/citation.cfm?id=1356052.1356053&coll=DL&dl=GUIDE&CFID=71223967&CFTOKEN=96440140}{Anatomy of high-performance matrix multiplication.}
	ACM Transactions on Mathematical Software (TOMS).
	Volume 34 Issue 3, May 2008, Article No. 12.
	Also available from 
	 \myhref{http://shpc.ices.utexas.edu/publications.html}{http://shpc.ices.utexas.edu/publications.html}.
\end{quote}
A further ``refactoring'' of this approach was more recently described in~\cite{BLIS1} 
\begin{quote}
	Field G. Van Zee, Robert A. van de Geijn. 
	\myhref{http://dl.acm.org/citation.cfm?id=2786970.2764454&coll=DL&dl=GUIDE&CFID=702354034&CFTOKEN=48470379}{%
		BLIS: A Framework for Rapidly Instantiating BLAS Functionality.} 
	ACM Transactions on Mathematical Software (TOMS).
	Volume 41 Issue 3, June 2015,
	Article No. 14. 
	Also available from \myhref{http://shpc.ices.utexas.edu/publications.html}{http://shpc.ices.utexas.edu/publications.html}.
\end{quote}
The advantage of the BLIS framework is that it reduces the kernel that must be highly optimized, possibly with vector intrinsics or in assemply code, to a {\em micro-kernel}.   In this section, we briefly describe the highlights of the approach.  However, we strongly suggest the reader become familiar with the above two papers themselves.

Figure~\ref{fig:blis_gemm}~(left) illustrates the way the Goto approach structures the blocking for three layers of cache (L1, L2, and L3).  In the BLIS framework, the implementation is structured exactly this way so that only the micro-kernel at the bottom needs to be highly optimized and customized for a given architecture.  In the original GotoBLAS implementation, now maintained as OpenBLAS~\cite{OpenBLASweb}, the operation starting with the second loop around the micro-kernel is instead customized.  In order to get the best performance, it helps is all data is accessed contiguously, which is why at some point prior to reaching the micro-kernel, data is packed in the order indicated by the arrows:
\begin{center}
	\includegraphics[width=2in]{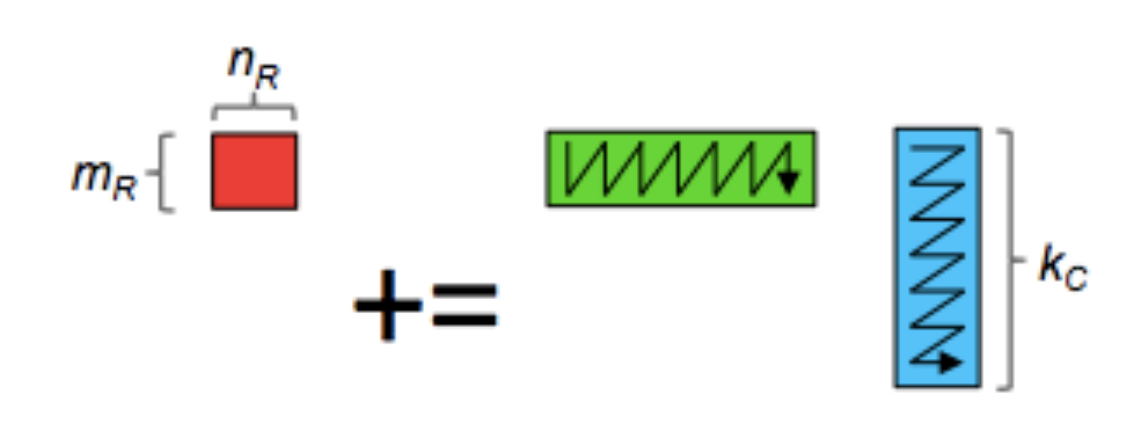}
\end{center}

Now, notice that each column of the block of $ A $ in the above picture is multiplied by each element in the corresponding row of the block of $ B $.  (We call these blocks of $ A $ and $ B $ {\em micro-panels}.)  This means that the {\em latency} to the L2 cache (the time required to bring in an element of the micro-panel of $ A $ from that cache) can be amortized over $ 2n_R $ flops.  For this reason, we can organize the computation so that the micro-panel of $ A $ typically resides in the L2 cache.  Actually, we can do better: while a rank-1 update is happening with a column of the micro-panels of $ A $ and $ B $, the next column of the micro-panel of $ A $ can be brought into registers so that computation masks the cost of that data movement.
The fact that we want to keep the micro-panel of $ B $ in the L1 cache (because it will be reused for many micro-panels of $ A $) limits the blocking parameter $ k_C $.

With the insights, the rest of the picture hopefully becomes clear.
The first loop around the microkernel works with a block of $ A $, $ \widetilde A_i $, that has been packed and resides in the L2 cache (by virtue of how the computation is ordered).  This limits the blocking parameter $ m_C $.  That block of $ A $ multiplies a block of $ B $, $ \widetilde B_p $, that has been packed to reside in the L3 cache (if the processor has an L3 cache).  Notice that the packing into  $ \widetilde A_i $ is amortized over all computation  with $ \widetilde B_p $ and the packing into $ \widetilde B_p $ is amortized over computations with many blocks $ A_i $.  The outermost loop partitions $ B $ so that the block $ \widetilde B_p $ fits in the L3 cache or, if a processor does not have an L3 cache, limits the amount of workspace for packing $ \widetilde B_p $ that is needed.  This limits the blocking parameter $ n_C $.

One may ask if the above described scheme is optimal.  In~\cite{ITXGEMM:ICCS01} a theory is given that shows that under an idealized model the above is locally optimal (in the sense that assuming data is in a certain memory layer in the hierarchy, the proposed blocking at that level optimally amortizes the cost of data movement with the next memory layer).  A theory that guides the choice of the various blocking parameters is given in~\cite{BLIS4}.

\subsection{Setup}

\begin{figure}[tb!]
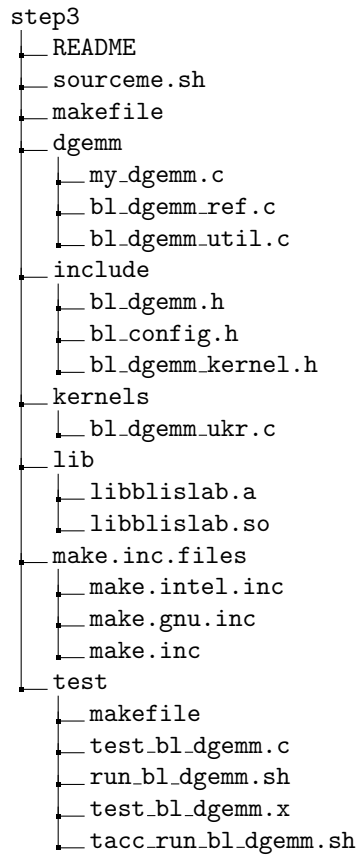

	\begin{center}
\begin{minipage}{4in}
	\dirtree{%
		.1 step3.
		.2 README 
.
		.2 sourceme.sh 
.
		.2 makefile 
.
		.2 dgemm 
.
		.3 my\_dgemm.c 
.
		.3 bl\_dgemm\_ref.c 
.
		.3 bl\_dgemm\_util.c 
.
		.2 include 
.
		.3 bl\_dgemm.h 
.
		.3 bl\_config.h 
.
		.3 bl\_dgemm\_kernel.h 
.
		.2 kernels
.
		.3 bl\_dgemm\_ukr.c
.
		.2 lib 
.
        .3 libblislab.a
.
        .3 libblislab.so
.
		.2 make.inc.files
.
		.3 make.intel.inc 
.
		.3 make.gnu.inc 
.
		.3 make.inc 
.
		.2 test 
.
		.3 makefile 
.
		.3 test\_bl\_dgemm.c 
.
		.3 run\_bl\_dgemm.sh 
.
		.3 test\_bl\_dgemm.x 
.       
		.3 tacc\_run\_bl\_dgemm.sh 
.	
	}
\end{minipage}
\end{center}
\caption{Structure of directory {\tt step3}.}
\label{fig:DirStep1}
\end{figure}

Figure~\ref{fig:DirStep1} illustrates the directory
structure for subdirectory {\tt step3}. Comparing to {\tt step1}, we have modified/added the following directories/files:

\begin{description}
\item[{\tt kernels}] This directory contains the micro-kernel implementations for various architecture.
\begin{description}
\item[{\tt bl\_dgemm\_ukr.c}] gives a naive C implementation.
\item[{\tt bl\_dgemm\_int\_kernel.c}] gives an {\tt AVX/AVX2} intrinsics micro-kernel implementation for Haswell architecture.
\item[{\tt bl\_dgemm\_asm\_kernel.c}] gives an {\tt AVX/AVX2} assembly micro-kernel implementation for Haswell architecture.
\end{description}
\end{description}

\subsection{Advanced techniques}
You can find the vector instructions online:
\begin{itemize}
\item \myhref{https://software.intel.com/sites/landingpage/IntrinsicsGuide/}{Intel Intrinsics Guide}
\item \myhref{https://software.intel.com/en-us/isa-extensions}{Intel ISA Extensions}
\end{itemize}

\subsubsection{An introduction example for ``{\tt axpy}''}
We provide you an example for the implementation of ``{\tt axpy}'' to demostrate how to use Intel {\tt AVX} Intrinsics and Assembly (in {\tt misc/examples} subdirectory). This example will serve as a good start point for you to learn basic {\tt broacast}/{\tt fma}/{\tt load}/{\tt store} instructions. Moreover, this example is actually a primitive for the ``broadcast'' implementation for 4$\times$4 rank-1 update.

\subsubsection{4$\times$4 rank-1 update}
\begin{figure}
\begin{center}
	\includegraphics[width=5in]{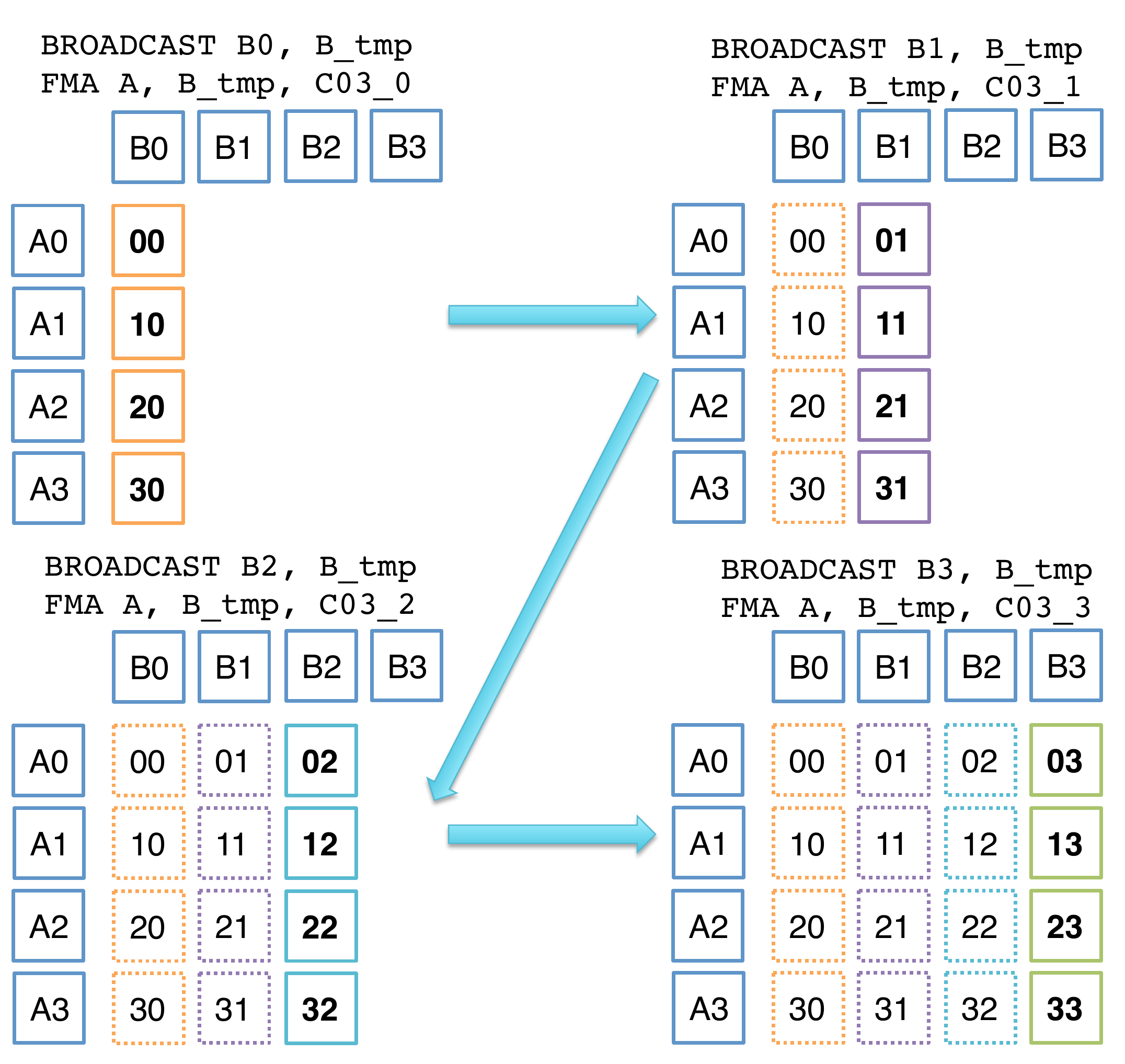}
\end{center}
\caption{{\tt AVX} 4$\times$4 rank-1 update with broadcast. Given 4$\times$1 vector $A$ and $B$, we compute the 4$\times$4 outer-product C by 4 {\tt FMA} interleaved with vectorized broadcast operations.}
\label{fig:broadcast}
\end{figure}

\begin{figure}[tb!]
\begin{center}
	\includegraphics[width=5in]{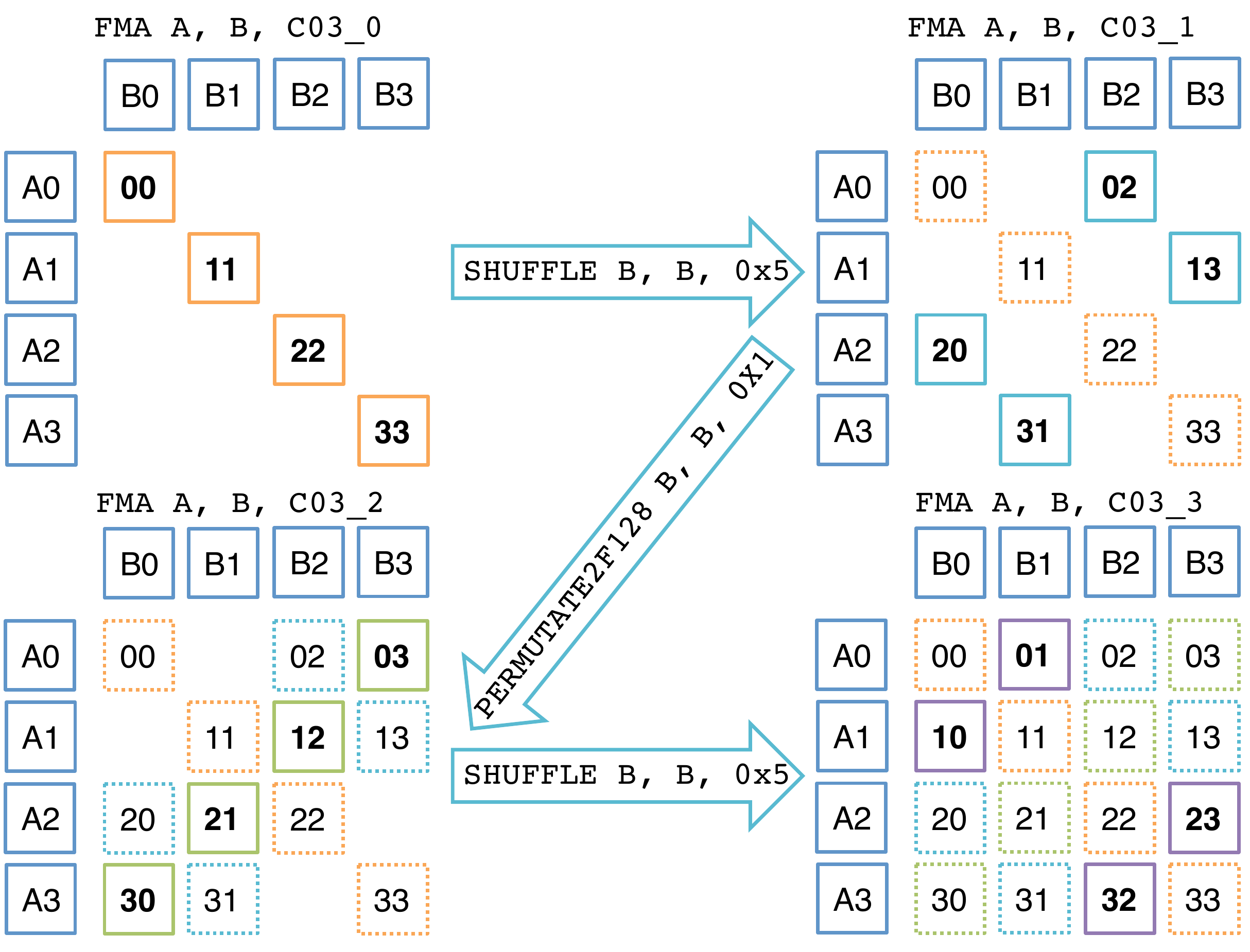}
\end{center}
\caption{{\tt AVX} 4$\times$4 rank-1 update with butterfly permutation. Given 4$\times$1 vector $A$ and $B$, we compute the 4$\times$4 outer-product C by 4 {\tt FMA} interleaved with vectorized shuffling operations. The 3rd operands (0x5, 0x1) indicates the shuffling (permutation) type.}
\label{fig:permutation}
\end{figure}

The micro-kernel implementation can be boiled down to a 4$\times$4 rank-1 update. There are two possible implementation: one based on broadcast (Figure~\ref{fig:broadcast}) and and one of a butterfly permutation (Figure~\ref{fig:permutation}). You can also try other possible implementations.

\subsection{Your mission, if you choose to accept it}
We provide you a reference implementation of simplified BLIS framework in {\tt my\_dgemm}. The code is organized in the same way presented in Figure~\ref{fig:blis_gemm}. However, the step size in each loop is not well choosen, and the micro-kernel implementation is a naive {\tt C} version. Therefore. you will not expect high performance with the code. What we want you to do is to 
\begin{itemize}
\item Specify the blocking parameter $ m_C $, $ n_C $, $ k_C $ and the micro-kernel size parameter $ m_R $, $ n_R $ in the file {\tt include/bl\_config.h}; and
\item Implement the efficient micro-kernel with vector intrinsics or assembly code. Place the code in {\tt kernels/bl\_dgemm\_int\_kernel.c} (for vector intrinsics), or {\tt kernels/bl\_dgemm\_asm\_kernel.c} (for assembly). You need to specify the function name of the micro-kernel by modifying {\tt BL\_MICRO\_KERNEL} in {\tt include/bl\_config.h}.
\end{itemize}

\section{Step 4: Parallelizing with OpenMP}

The benefit of the BLIS way of structuring the GotoBLAS approach to the implementation of \Gemm\ is that it exposes five loops in {tt C} which can then be easily parallelized with OpenMP directives.

\subsection{To parallelize or not to parallelize, that's the question}

The fundamental question becomes which loop to parallelize.  In~\cite{BLIS3}
\begin{quote}
	Tyler M. Smith, Robert van de Geijn, Mikhail  Smelyanskiy, Jeff R. Hammond, and Field G. Van Zee.
	\myhref{http://ieeexplore.ieee.org/xpl/articleDetails.jsp?arnumber=6877334}{Anatomy of High-Performance Many-Threaded Matrix Multiplication.} IEEE 28th International Parallel and Distributed Processing Symposium, 2014.  Also available from \myhref{http://shpc.ices.utexas.edu/publications.html}{http://shpc.ices.utexas.edu/publications.html}.
\end{quote} 
a detailed discussion is given of what the pros and cons are regarding the parallelization of each loop.
For multi-core architectures (multi-threaded architectures with relatively few cores) results can be found in the earlier paper~\cite{BLIS2} 
\begin{quote}
	Field G. Van Zee, Tyler Smith, Bryan Marker, Tze Meng Low, Robert A. van de Geijn, Francisco D. Igual, Mikhail Smelyanskiy, Xianyi Zhang, Michael Kistler, Vernon Austel, John Gunnels, Lee Killough. 
	\myhref{}{The BLIS Framework: Experiments in Portability.}		
		ACM Transactions on Mathematical Software.  To appear. Also available from \myhref{http://shpc.ices.utexas.edu/publications.html}{http://shpc.ices.utexas.edu/publications.html}.
\end{quote}

\section{Conclusion}
\label{sec:conclusion}

We use GEMM as a case study to show how to program for performance.

\subsection*{Useful links}

\noindent
Documentation
\begin{itemize}
	\item 
    \myhref{http://shpc.ices.utexas.edu/}{The Science of High-Performance Computing (SHPC) group website}.
    \item
    \myhref{http://www.cs.utexas.edu/users/flame/web/FLAMEPublications.html}{The FLAME project publications webpage}.  (The umbrella project that BLIS is part of is known as the FLAME project.)
    \item \myhref{https://software.intel.com/sites/landingpage/IntrinsicsGuide/}{Intel Intrinsics Guide}.
    \item \myhref{https://software.intel.com/en-us/isa-extensions}{Intel ISA Extensions}.
    \end{itemize}
    
\noindent
Software
\begin{itemize}
	\item
	\myhref{http://https://github.com/flame/blis}{BLIS on GitHub}.
	\item
	\myhref{https://software.intel.com/en-us/qualify-for-free-software}{Intel Free Software} (including C/C++ compilers).
	\item
	\myhref{https://software.intel.com/en-us/intel-mkl}{Intel's Math Kernels Library (MKL) website}.
	\item
	\myhref{https://software.intel.com/en-us/articles/free_mkl}{Download MKL for free}.
\end{itemize}

\subsection*{Acknowledgments}
We thank the other members of the Science of High-Performance Computing (SHPC) team for their support.
This research was partially sponsored by the Nantional Science Foundation grant ACI-1148125/1340293 and Intel through its funding of the SHPC group as an Intel Parallel Computing Center.

{\em Any opinions, findings and conclusions or recommendations
expressed in this material are those of the author(s) and do not
necessarily reflect the views of the National Science Foundation
(NSF).}

\bibliographystyle{plain} 
\bibliography{biblio}

\begin{thebibliography}{10}

\bibitem{IBM:P2}
R.C. Agarwal, F.G. Gustavson, and M.~Zubair.
\newblock Exploiting functional parallelism of {POWER2} to design
  high-performance numerical algorithms.
\newblock {\em {IBM} {J}ournal of {R}esearch and {D}evelopment}, 38(5), Sept.
  1994.

\bibitem{LAPACK3}
E.~Anderson, Z.~Bai, C.~Bischof, L.~S. Blackford, J.~Demmel, Jack~J. Dongarra,
  J.~Du Croz, S.~Hammarling, A.~Greenbaum, A.~McKenney, and D.~Sorensen.
\newblock {\em LAPACK Users' guide (third ed.)}.
\newblock Society for Industrial and Applied Mathematics, Philadelphia, PA,
  USA, 1999.

\bibitem{PHiPAC97}
Jeff Bilmes, Krste Asanovi\'c, Chee whye Chin, and Jim Demmel.
\newblock Optimizing matrix multiply using \mbox{PHiPAC}: a {P}ortable,
  {H}igh-{P}erformance, \mbox{ANSI} {C} coding methodology.
\newblock In {\em Proceedings of International Conference on Supercomputing},
  Vienna, Austria, July 1997.

\bibitem{BLAS3}
Jack~J. Dongarra, Jeremy Du~Croz, Sven Hammarling, and Iain Duff.
\newblock A set of level 3 basic linear algebra subprograms.
\newblock {\em ACM Trans. Math. Soft.}, 16(1):1--17, March 1990.

\bibitem{BLAS2}
Jack~J. Dongarra, Jeremy Du~Croz, Sven Hammarling, and Richard~J. Hanson.
\newblock An extended set of {FORTRAN} basic linear algebra subprograms.
\newblock {\em ACM Trans. Math. Soft.}, 14(1):1--17, March 1988.

\bibitem{Goto:2008:AHP}
Kazushige Goto and Robert~A. van~de Geijn.
\newblock Anatomy of a high-performance matrix multiplication.
\newblock {\em ACM Trans. Math. Soft.}, 34(3):12, May 2008.
\newblock Article 12, 25 pages, Available from
  \myhref{http://shpc.ices.utexas.edu/publications.html}{http://shpc.ices.utexas.edu/publications.html}.

\bibitem{ITXGEMM:ICCS01}
John~A. Gunnels, Greg~M. Henry, and Robert~A. van~de Geijn.
\newblock A family of high-performance matrix multiplication algorithms.
\newblock In Vassil~N. Alexandrov, Jack~J. Dongarra, Benjoe~A. Juliano,
  Ren\'e~S. Renner, and C.J.~Kenneth Tan, editors, {\em {C}omputational
  {S}cience - {ICCS 2001}, {P}art {I}}, Lecture Notes in Computer Science 2073,
  pages 51--60. Springer-Verlag, 2001.

\bibitem{BLIS-TI}
Francisco~D. Igual, Murtaza Ali, Arnon Friedmann, Eric Stotzer, Timothy Wentz,
  , and Robert {v}an~{d}e {G}eijn.
\newblock Unleashing the high-performance and low-power of multi-core dsps for
  general-purpose hpc.
\newblock In {\em SC12}, 2012.

\bibitem{poorman_journal}
B.~K{\aa}gstr\"{o}m, P.~Ling, and C.~Van Loan.
\newblock {GEMM}-based level 3 {BLAS}: High performance model implementations
  and performance evaluation benchmark.
\newblock {\em ACM Trans. Math. Soft.}, 24(3):268--302, 1998.

\bibitem{BLAS1}
C.~L. Lawson, R.~J. Hanson, D.~R. Kincaid, and F.~T. Krogh.
\newblock Basic linear algebra subprograms for {F}ortran usage.
\newblock {\em ACM Trans. Math. Soft.}, 5(3):308--323, Sept. 1979.

\bibitem{OpenBLASweb}
Open{BLAS}, an optimized {BLAS} library.
\newblock \url{http://www.openblas.net}.

\bibitem{BLIS3}
Tyler~M. Smith, Robert {v}an~{d}e {G}eijn, Mikhail Smelyanskiy, Jeff~R.
  Hammond, , and Field~G. {V}an {Z}ee.
\newblock Anatomy of high-performance many-threaded matrix multiplication.
\newblock In {\em International Parallel and Distributed Processing Symposium
  2014}, 2014.

\bibitem{BLIS4}
{T}ze~{M}eng Low, Francisco~D. Igual, Tyler~M. Smith, and Enrique~S.
  Quintana-Ort\'{\i}.
\newblock Analytical modeling is enough for high performance blis.
\newblock {\em {ACM} Transactions on Mathematical Software}.
\newblock in review. Available from
  \myhref{http://shpc.ices.utexas.edu/publications.html}{http://shpc.ices.utexas.edu/publications.html}.

\bibitem{BLIS-Encycl}
Robert {v}an~{d}e {G}eijn and Kazushige Goto.
\newblock {\em Encyclopedia of Parallel Computing, Part 2}, chapter Robert van
  de Geijn and Kazushige Goto, pages 157--164.
\newblock Springer, 2011.

\bibitem{BLIS2}
Field~G. {V}an {Z}ee, Tyler Smith, Bryan Marker, Tze~Meng Low, Robert~A.
  {v}an~{d}e {G}eijn, Francisco~D. Igual, Mikhail Smelyanskiy, Xianyi Zhang,
  Michael Kistler, Vernon Austel, John Gunnels, and Lee Killough.
\newblock The blis framework: Experiments in portability.
\newblock {\em {ACM} Transactions on Mathematical Software}.
\newblock to appear.

\bibitem{BLIS1}
Field~G. {V}an {Z}ee and Robert~A. {v}an~{d}e {G}eijn.
\newblock {BLIS}: A framework for rapidly instantiating blas functionality
  (replicated computational results certified).
\newblock {\em ACM Trans. Math. Soft.}, 41(3):14:1--14:33, June 2015.
\newblock Available from
  \myhref{http://shpc.ices.utexas.edu/publications.html}{http://shpc.ices.utexas.edu/publications.html}.

\bibitem{ATLAS}
R.~Clint Whaley and Jack~J. Dongarra.
\newblock Automatically tuned linear algebra software.
\newblock In {\em Proceedings of SC'98}, 1998.

\bibitem{ATLAS_journal}
R.~Clint Whaley, Antoine Petitet, and Jack~J. Dongarra.
\newblock Automated empirical optimizations of software and the atlas project.
\newblock {\em Parallel Computing}, 27(1-2):3--35, 2001.

\bibitem{libflame_ref}
Field G.~Van Zee.
\newblock {\em {\tt libflame}: {T}he {C}omplete {R}eference}.
\newblock {\tt www.lulu.com}, 2009.

\end{thebibliography}

\end{document}